\documentclass[aps,pre,twocolumn,superscriptaddress]{revtex4}

\usepackage{amsfonts,amssymb,amsmath,latexsym,epsfig,mathrsfs}
\usepackage{graphicx}
\usepackage{color}
\usepackage{url} 
\newcommand{\supp}{{\rm supp\,}}

\begin{document}              

\title{Escape from attracting sets in randomly perturbed systems \\ Phys. Rev. E 82, 046217 (2010) }

\author{Christian
S. Rodrigues}~\email{christian.rodrigues@mis.mpg.de}

\affiliation{Max Planck Institute for Mathematics in the Sciences, Inselstr., 22, 04103 Leipzig, Germany}
\affiliation{Department of Physics and Institute for Complex Systems and Mathematical Biology,
King's College, University of Aberdeen - Aberdeen AB24 3UE, UK}

\author{Celso Grebogi} 
\affiliation{Department of Physics and Institute for Complex Systems and Mathematical Biology,
  King's College, University of Aberdeen - Aberdeen AB24 3UE, UK}

\author{Alessandro P. S. de Moura}
\affiliation{Department of Physics and Institute for Complex Systems and Mathematical Biology,
King's College, University of Aberdeen - Aberdeen AB24 3UE, UK}

\date{\today}

\begin{abstract}


The dynamics of escape from an attractive state due to random
perturbations is of central interest to many areas in science.
Previous studies of escape in chaotic systems have rather focused on
the case of unbounded noise, usually assumed to have Gaussian
distribution. In this paper, we address the problem of escape induced
by bounded noise. We show that the dynamics of escape from an
attractor's basin is equivalent to that of a closed system with an
appropriately chosen ``hole''. Using this
equivalence, we show that there is a minimum noise amplitude above
which escape takes place, and we derive analytical expressions for the
scaling of the escape rate with noise amplitude near the escape
transition. We verify our analytical predictions through numerical
simulations of two well known $2$-dimensional maps with noise.
\end{abstract}


\maketitle


\indent

The escape of trajectories from attracting sets due to the effect of
noise has been a central issue in various branches of science for a
long time. From a fundamental perspective, the study of the dynamics of escape includes the
fundamental work by Arrhenius on chemical reactions~\cite{Han86},
passing through ideas of Kramers from the forties of last
century~\cite{Kra40}, the escaping on chaotic dynamics~\cite{Gra89,
  KFG99, KrF02, KrG04} to very recent work on Statistical
Mechanics~\cite{DeG09}. Notwithstanding this long history, an
important case has mostly been neglected, namely the dynamics under
bounded noise. In fact, the vast majority of theoretical works in this area is heavily dependent on the assumptions of 
unbounded noise, almost always assumed to have a Gaussian
distribution~\cite{Gra89, Bea89, DeG09}. 
Thus, they are not applicable to other cases. In particular, very little is known about the dynamics of
systems with escape in the presence of bounded noise.  As for many
applications a bounded perturbation is arguably more realistic than
unbounded ones, this is an important gap in our understanding.
For example, in Neuroscience one may be interested in the
minimum energy for bursting to take place~\cite{NNM00}, in Geophysics
one may consider the overcoming of some potential barrier just before
an earthquake~\cite{PeC02}, or the critical outbreak magnitude for the
spread of epidemics~\cite{BBS02}, among others. 

In this paper, we investigate escape in dynamics perturbed by bounded
noise using a new approach. We describe the noisy dynamics in terms of
a family of random maps, whose iteration gives rise to a discrete
Markov process with transition probability supported in the
neighbourhood of the points generated by the iteration of the
deterministic dynamics. The dynamics near an attractor is then
described using the formalism of conditionally invariant measures. We
show that the dynamics of the escape is determined by the measure of a
subset $I_{\partial}$ of the phase space, which we call
\textit{conditional boundary}. It consists of those points lying close
enough to the basin boundary such that they are subject to escape
under the effect of random perturbation in one iteration of the map,
and intersecting the support of the conditionally invariant measure of
the system. Therefore, the dynamics is mapped onto the dynamics of a
closed system with a hole --- $I_{\partial}$ being the hole.  We show
that there is a minimum noise amplitude for the escape to take place,
which is the critical noise amplitude $\xi_{c}$ that makes
$I_{\partial}$ non-empty. We show that the mean escape time is
determined by the measure of the conditional boundary, and use this to
derive a power law relation between the average escape time $\langle T
\rangle$ and the noise amplitude $\xi$, for $\xi$ close to (and higher
than) $\xi_c$:
\begin{equation}
\label{eq.meantime}
\langle T \rangle \approx (\xi - \xi_{c})^{-\alpha},
\end{equation}
where $\alpha$ depends on the dimension of the system. We show that
for dimension two, $\alpha = 3/2$, and we verify that this prediction
is correct by comparing with the results of numerical
simulations for two maps from different families. This result is independent of any particular system, and
holds universally for bounded noise. Equation (\ref{eq.meantime}) is in
contrast with the case of Gaussian noise, where an exponential scaling
is observed~\cite{Han86, Kra40, KFG99, KrF02, KrG04, DeG09, Gra89,
  Bea89}. Furthermore, although the noise-induced escape from
attractors may seem to be very different from that of systems
undergoing a boundary bifurcation, we show that our approach allows us
to establish a connection between these two processes, and to explain why
we find the same time scaling in both cases.

To get started, first consider a deterministic dynamics $x_{n+1} =
f(x_{n})$ given by the iteration of the map $f$, a smooth function whose inverse is differentiable, in the phase
space of the system $M$ (i.e., the iteration of a diffeomorphism $f: M \rightarrow M$). Our main focus will be on invariant subsets $\Lambda$ of
$M$ which attract their neighbouring points, that is, $f^{n}(x)$ tends
to $\Lambda$ as $n \rightarrow \infty$; these are the attractors of the
system. The
basin of attraction of $\Lambda$ is the open set $W^{s}(\Lambda)$, the set of
points eventually coming close and converging to
$\Lambda$. The next ingredient is the boundary of the basin of
attraction, a zero Lebesgue measure ergodic component of the phase
space which we denote by $\partial$.

\begin{figure}[tb]
\includegraphics[width=.65\columnwidth]{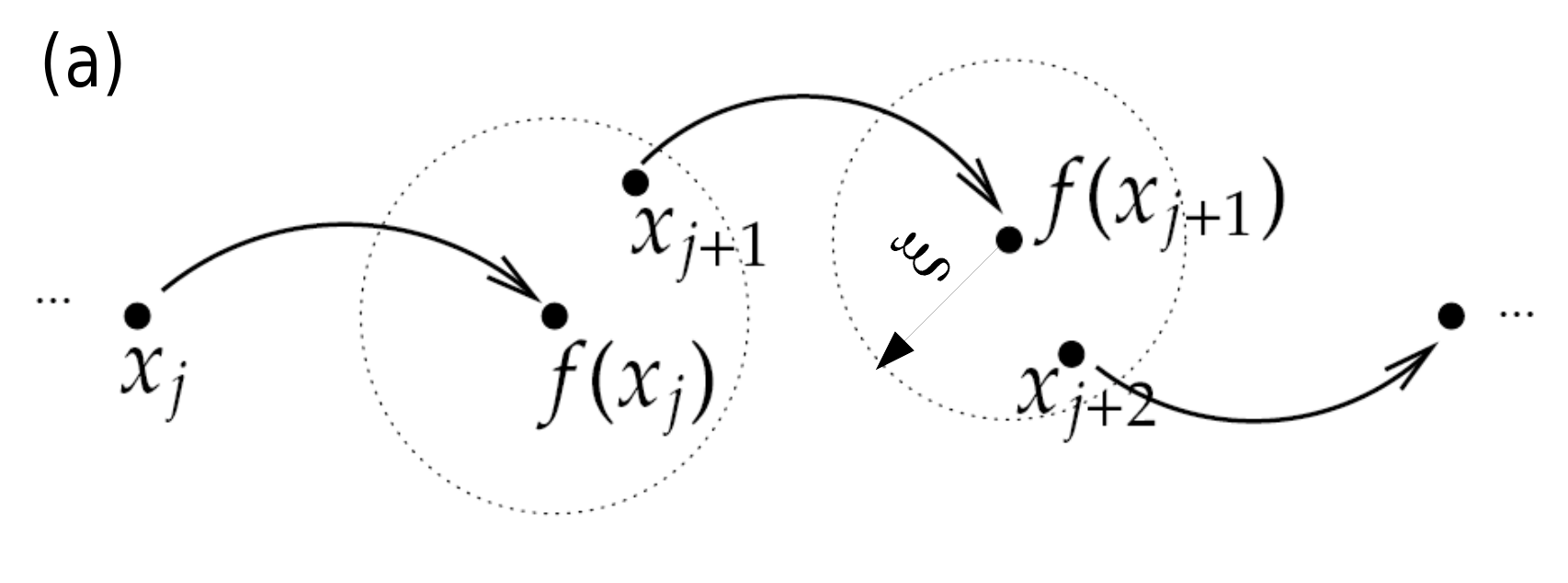}\\
\vspace{0.5cm}
\includegraphics[width=.65\columnwidth]{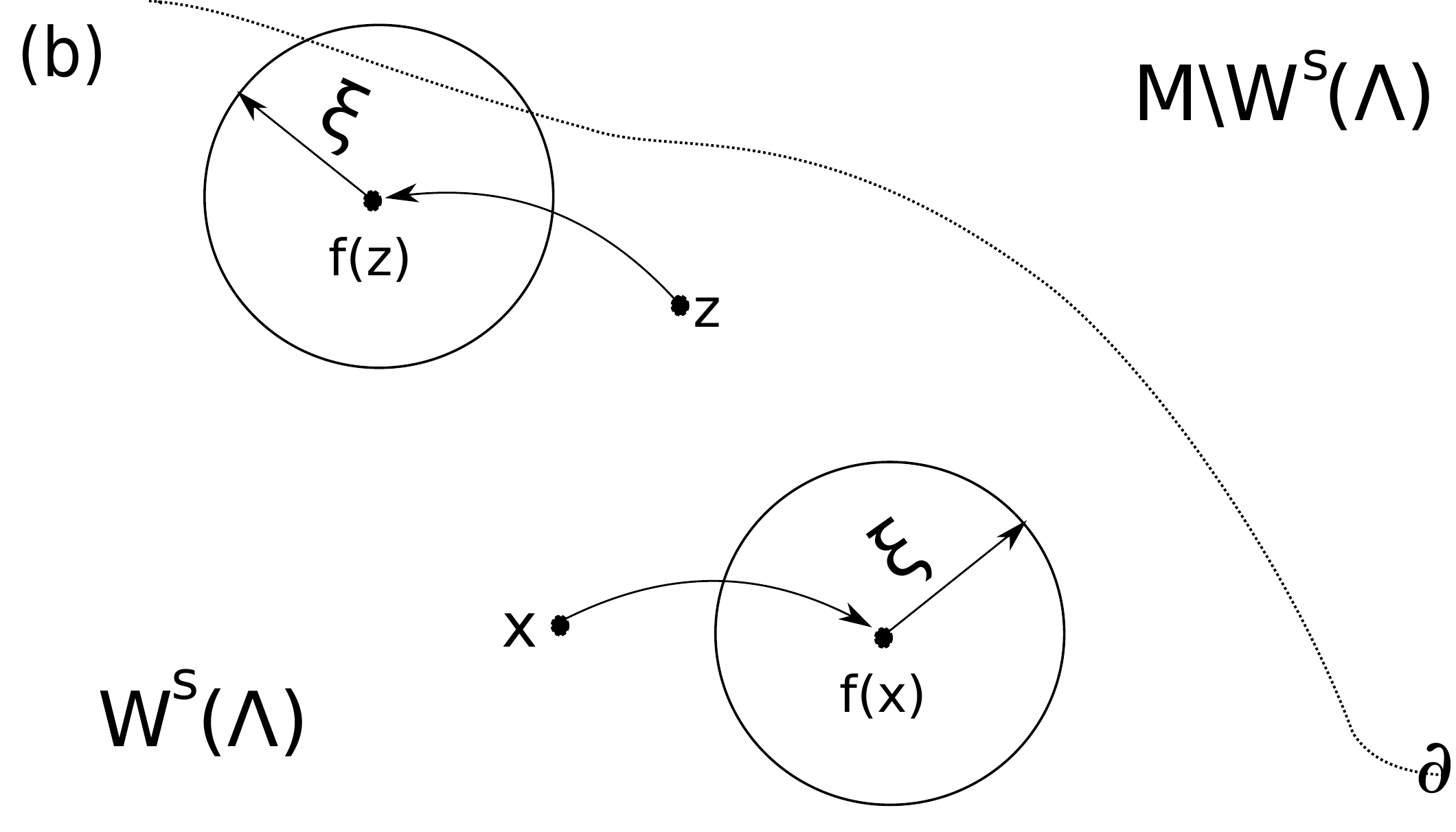}
\caption{(a) Illustrative picture of the perturbed dynamics with amplitude
  of noise $||\varepsilon_{j}|| < \xi$. (b) A basin of attraction $W^{s}(\Lambda)$
  and its basin boundary $\partial$ (dashed line). We illustrate that the iteration $z \mapsto f(z)$,
  from the point $z$ initially in $W^{s}(\Lambda)$ brings the
  orbit within a distance $\xi$ from the boundary. Therefore, the
  random perturbation applied to $f(z)$ with some $||\varepsilon|| <
  \xi$ could push the random orbit outside the basin. On the other hand, for the point $x \in
  W^{s}(\Lambda)$ the iteration $x \mapsto f(x)$ brings it farther
  than the maximum perturbation $\xi$ away from the boundary
  $\partial$. Therefore, in our illustration $z \in
  \tilde{I_{\partial}}$ but $x \notin \tilde{I_{\partial}}$.}
\label{fig.function}
\end{figure}

We shall consider a random perturbation of the deterministic
system introduced above~\cite{Ara00, BDV05}, so that our perturbed
system is described by a family of \textit{random maps}, that in our
context can be written as
\begin{equation}
F(x_{j}) = f(x_{j}) + \varepsilon_{j},
\end{equation}
with $||\varepsilon_{j}|| < \xi$, where $\varepsilon_{j}$ is the
vector of random noise added to the
deterministic dynamics at the iteration $j$, and $\xi$ is its maximum
amplitude. In this way, $F$ is a continuous application~\footnote{The iteration of random maps can
also be seen as an associated discrete time Markov process having a
family $\{p_{\xi}(\cdot | x) : x \in U, \xi >0\}$ of transition
probabilities, where every $p_{\xi}(\cdot | x)$ is supported on $U$, a
$\xi$-neighbourhood of $f(x)$.}. We illustrate
this in Fig.~\ref{fig.function}(a). As it is shown in the picture, the perturbed dynamics can be thought as follow. Image we iterate the point $x_{j}$ by the deterministic system $f$. Then, let say that at the moment that we take the $f(x_{j})$ to evolve our dynamic again, we make a small error given to our limited precision. Nevertheless, we can assure that the error is always less than $\xi$. Then we ask whether the attractors for the system with no error are still attractors when a small error is considered. If so, how large can our error be such as the attractors will still be preserved? Above this threshold, how does the escape of orbits scale? These natural extensions of deterministic processes are exactly the sort of problems we shall be interested in. 

 From the probabilistic perspective, the idea of orbits converging to
 some attractor is represented by the concept of \textit{physical} (or
 SRB) measures~\cite{BDV05, EcR85}. Suppose initially that we compute the time average, $\frac{1}{n} \sum_{j=0}^{n-1} f^{j}(x)$,  along the orbit for a given initial condition as $n$ evolves. This quantity, the time average, is expected to converge to some invariant value as the orbit approaches the attractor. If we extend it for a large number of initial conditions, then, the time averages that we compute along different orbits are also expected to converge to an invariant value as these orbits approach the attractor. It turns out that such value defines the so-called SRB measure, which characterises the attractor. The set of initial conditions that we chose for computing such time averages along the orbits are exactly what we call the \textit{basin of the measure}, that we represent by $\mathscr{B}(\mu)$. Since we want to statistically characterise the behaviour of the attractor, it is desirable the set of initial conditions for which we compute such time averages to have positive Lebesgue measure. That is, there is a non negligible number of initial conditions whose time averages along the orbits converge to such invariant quantity and thus characterise the attractor. Therefore, we say that a measure is physical if its basin has positive Lebesgue measure.
More precisely, the basin $\mathscr{B}(\mu)$ of
 such measure is the set of points whose time averages along the
 orbits weakly converge to the space average, $\lim_{n \rightarrow
   \infty} \frac{1}{n} \sum_{j=0}^{n-1}\varphi (f^{j}(x)) = \int
 \varphi d\mu$, where $\varphi$ represents our measurements of an observable~\footnote{More generally, for every continuous function $\varphi :M \rightarrow
 \mathbb{R}$.}. Here, a physical measure of a set $A
 \subset M$ is an invariant ergodic probability measure $\mu$
 supported on $A$. Therefore, such measures are closely related to the
 so-called \textit{natural} measures. Note that as a consequence of
 the Birkhoff's ergodic theorem,
\begin{equation}
\label{eq.basmeasur}
\mu(\mathscr{B}(\mu)) = 1.
\end{equation}
In words, Eq.~\ref{eq.basmeasur} just says that we are normalising such measure on the total set of initial conditions whose time averages converge to the invariant one. For most known cases the basin of the
physical measures supported on the attractors coincides with the basins
of the attractors~\cite{BDV05}, so we assume here that,
$W^{s}(\Lambda) = \mathscr{B}(\mu)$.
This is known as the \emph{basin property}.

Despite the complications introduced by the presence of noise in the
dynamics, for continuous random maps with small amplitude of
perturbation an invariant probability measure is guaranteed to exist. This is due to the
\textit{Krylov - Bogolubov Theorem}~\cite{Man87}, which ensures
that every continuous application in a compact measurable space has
an invariant probability measure. For randomly perturbed systems, we expect that
the perturbed orbit will densely fill up the neighbourhood of the attractor. This property is called \textit{random
  transitivity}~\cite{Ara00}, and under general conditions it can be
shown that these measures are ergodic~\cite{Ara00}. As an example, we
can think of an attractive fixed point for a deterministic system, and
the perturbed version of that system. In the perturbed dynamics, there
is a density of probability around the fixed point.  When we increase
the amplitude of the noise, the density of probability becomes more
spread around the original fixed point. Conversely, decreasing the
amplitude of noise, the support of the invariant measure tends to the
point attractor.

Now consider a subset $\tilde{I_{\partial}}$ of the phase space
neighbouring the basin boundary, defined by the set of points $x$ whose
$f(x)$ is within a distance $\xi$ from some point in the boundary
$\partial$,
\begin{equation}
\label{eq.thickbou}
\tilde{I_{\partial}}= \{x \in M; B(f(x),\xi) \cap \partial \neq \emptyset \},
\end{equation}
where $B(f(x), \xi)$ is the ball of radio $\xi$ around $f(x)$. This
is illustrated in Fig.~\ref{fig.function}(b). In addition to being
close enough to the boundary, another important condition for the
escaping process to take place is that the intersection of the support
of the invariant physical measure of the system under stochastic
perturbation, $\supp \mu_{\xi}$, overlaps the set
$\tilde{I_{\partial}}$. That is, the density of probability around the attractor needs to be spread over a region close enough to the boundary. We thus define the set $I_\partial$ by
\begin{equation}
\label{eq.bousupp}
I_{\partial} = \tilde{I_{\partial}} \cap \supp \mu_{\xi}.
\end{equation}
Because the noise is bounded, for very small noise amplitudes
$I_\partial = \emptyset$.  But as the amplitude of the noise is
increased, we expect that for a certain critical amplitude
$\xi=\xi_{c}$, we have $I_\partial \neq \emptyset$,
and escape takes place for any $\xi > \xi_{c}$. We call $I_{\partial}$ the
\textbf{conditional boundary}.  The importance of $I_{\partial}(\xi)$ stems from
the fact that it represents the set of points which one
iteration of the map $f$ can potentially send close enough to the boundary
$\partial$, such that a random perturbation with amplitude $\xi$ may
send them out of the basin of attraction. The dynamics of escape is
thus governed by this set, and it can be understood as a ``hole''
which sucks orbits from the basin if they land on it.
\begin{figure}[tb]
\includegraphics[width=.45\columnwidth]{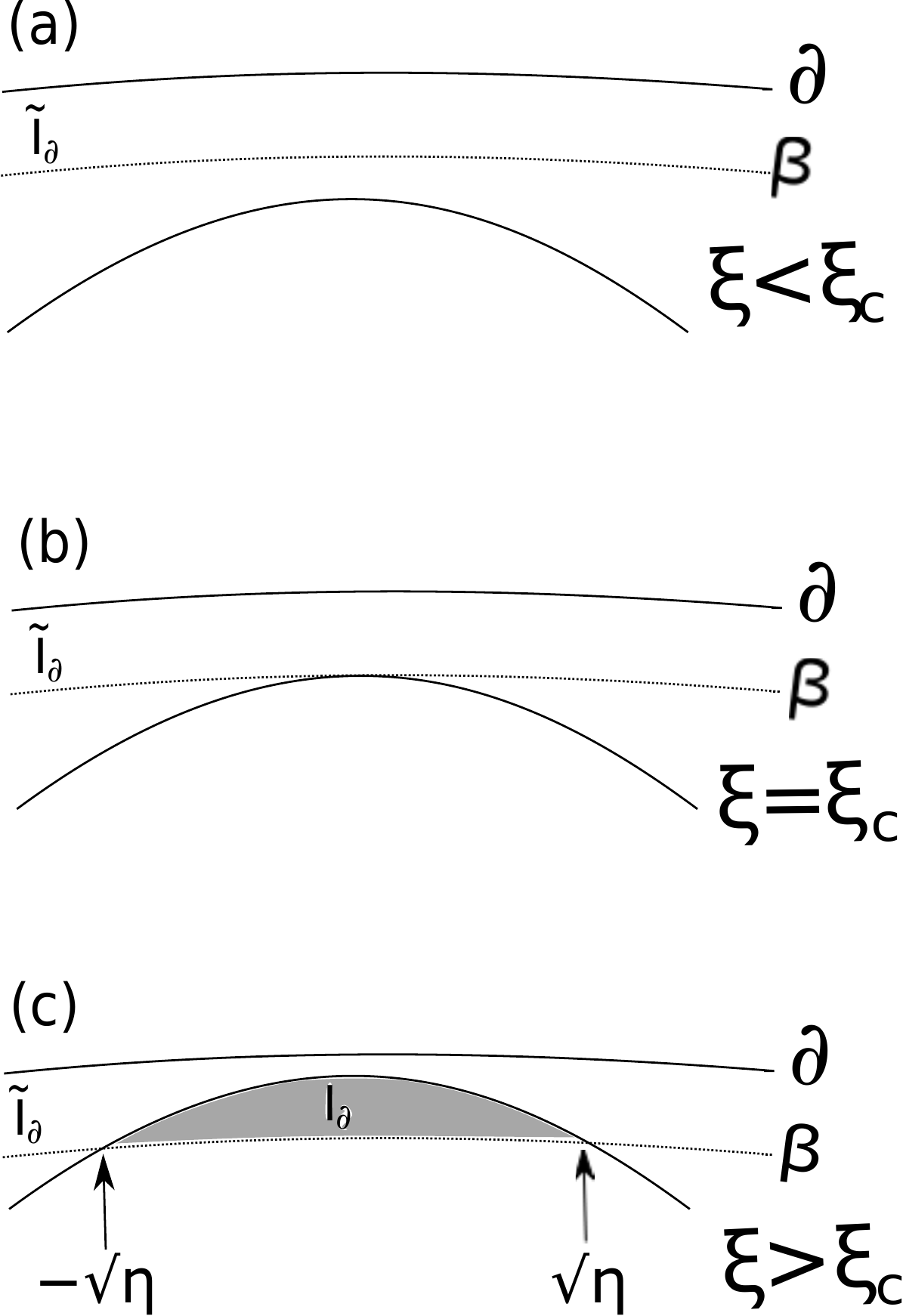}
\caption{We illustrate the basin boundary $\partial$, the region $\tilde{I_{\partial}}$ whose edge is the dashed line $\beta$, and the boundary of the support of the invariant measure for the perturbed system (the parabolic line). In (a), $\xi < \xi_{c}$, thus  $\tilde{I_{\partial}} \cap \supp \mu_{\xi} = \emptyset$. In (b), $\xi = \xi_{c}$ the $\supp \mu_{\xi}$ is tangent to the limit of $\tilde{I_{\partial}}$. In (c), $\xi > \xi_{c}$. As a consequence, $\tilde{I_{\partial}} \cap \supp \mu_{\xi} \neq \emptyset$, therefore, $I_{\partial} \neq \emptyset$ (shadowed area), what implies $\mu(I_{\partial}) >0$.}
\label{fig.condbound}
\end{figure}

Due to property (\ref{eq.basmeasur}), we have that $I_{\partial} \neq
\emptyset$ for $\xi>\xi_{c}$, and thus $ \mu(I_{\partial}) > 0$.
We can therefore think of the system with $\xi>\xi_c$ as a closed
system with a hole (or leak)~\cite{DeY06}, where $I_{\partial}$ plays
the role of the hole. The idea of a system with a hole has been used in different contexts before, for example~\cite{DeY06}. Furthermore, note that the measure
$\mu({I_{\partial}})$ is not in fact invariant because of the loss caused by
escape. However, it is possible to describe such escape problem in
terms of \textit{conditionally invariant measure}~\cite{ZmH07,
  DeY06}. We say that a probability measure $\mu_{c}$ is
\textit{conditionally invariant} with respect to $F$ if
\begin{equation}
\frac{\mu_{c}(F^{-1}(X))}{\mu_{c}(F^{-1}(A))} = \mu_{c}(X),
\end{equation}
for every measurable subset $X \subset A$, where $A \subset M$ is a
non-invariant region of the phase space $M$. The conditional measure
is defined in a way that, for each iteration, when the set $A$ loses a
fraction of its orbits to the hole, we renormalise its measure by what
remains in $A$. It is defined in terms of pre-images (or more
generally using \textit{pushforward} measures~\cite{DeY06}.) In
our context, the hole is the set $I_{\partial}$, and therefore
$\mu_{c}(I_{\partial})>0$ is preserved by the dynamics because of the
compensation factor $\mu_{c}(F^{-1}(A))$~\cite{DeY06, AlT08}. Due to
the random transitivity of the conditionally invariant domain, we
expect no particular dependence on the density of points in $A$. In
this case, a random trajectory diffuses through the support of
$\mu_{c}$, until it eventually comes inside $I_{\partial}$. Once in
this set, there is a probability that it will permanently
escape. Indeed it is the measure of $I_{\partial}$ which controls the
escape rate. Because we are interested in the regime of $\xi \gtrapprox
\xi_{c}$, hence small leaks, we can assume~\cite{AlT08},
\begin{equation}
\label{eq.cond}
\mu_{c}(I_{\partial}) = \mu(I_{\partial})>0. 
\end{equation}
Since there is a certain
probability that a particle escapes if it falls into $I_\partial$, we
expect from Kac's Lemma~\cite{Kak59} that the average escape time
satisfies
\begin{equation}
\label{eq.kac}
\langle T \rangle \propto \frac{1}{\mu(I_{\partial})}.
\end{equation}

Rigourously proving the existence of and calculating
$\mu(I_{\partial})$ is not an easy task. We use here a heuristic
approach to obtain the scaling of $\mu(I_\partial)$ for $\xi$ close to
$\xi_c$~\cite{LLB02}. For simplicity we focus on the $2$-dimensional
case. For $\xi < \xi_{c}$, we expect the probability distribution to be relatively concentrated on the attractor, thus having support with Lebesgue measure (area) smaller than that of the basin of attraction. As a first approximation, we image the edge of the support as being a smooth closed curve. As $\xi$ is increased, the mean radio of the distribution grows but it needs to be less than that of the basin of attraction, otherwise we would have the invariant measure supported outside the basin of attraction.  
When $\xi \gtrapprox \xi_{c}$, we picture it as a generic intersection of two curves of different radios.
Therefore, the portion of the edge of the support of the invariant measure lying within
$\tilde{I}_\partial$ is locally well approximated by a parabola (curved line
in Fig.~\ref{fig.condbound}).  For $\xi \gtrapprox \xi_{c}$, we have
$I_{\partial} \neq \emptyset$, thus $\mu_{c}(I_{\partial}) >0$. Recalling Eq.~\ref{eq.cond}, it gives us that $\mu(I_{\partial}) >0$.

To estimate the measure of the hole, define 
\begin{equation}
\eta = \xi - \xi_{c}. 
\end{equation}
The top curve encompassing the shadowed area
in Fig.~\ref{fig.condbound}(c) is then well approximated by 
\begin{equation}
S = \eta - \beta^{2},
\end{equation}
 in the appropriate units. $S$ intersects $\beta$ in two
points, namely $-\sqrt{\eta}$ and $\sqrt{\eta}$. Because we assumed the basin property to hold, in other words, we normalised the measure and assumed the basin of the measure to be equal to the basin of attraction of the deterministic system, Eq.~\ref{eq.basmeasur} also tells us that the probability measure of the hole is proportional to its Lebesgue measure, the area,
encompassed for $\beta$ and $S$ within the interval $[-\sqrt{\eta},
  \sqrt{\eta}]$ --- the shadowed area in Fig.~\ref{fig.condbound}(c).
We can calculate it as
\begin{equation}
\begin{split}
\mu(I_{\partial}) &\approx \int_{-\sqrt{\eta}}^{\sqrt{\eta}} S d\beta
= \frac{4}{3}\eta^{3/2} \quad \text{ , and thus}\\
\mu(I_{\partial}) &\approx (\xi - \xi_{c})^{3/2}.
\end{split}
\end{equation}
Applying the Kac's Lemma, we obtain Eq.~\ref{eq.meantime} with
$\alpha=3/2$. Note that although we use the approximation to describe the boundary $\partial$ locally as a smooth curve, it might actually be fractal. Therefore, if the random orbit falls into $I_{\partial}$, there is only a probability that it will escape due to the fractal property of $\partial$. This fact is subtly incorporated by Eq.~\ref{eq.kac} in the proportionality rather than the equality to the inverse of the mean escape time.
\begin{figure}[tb]
\includegraphics[width=1\columnwidth]{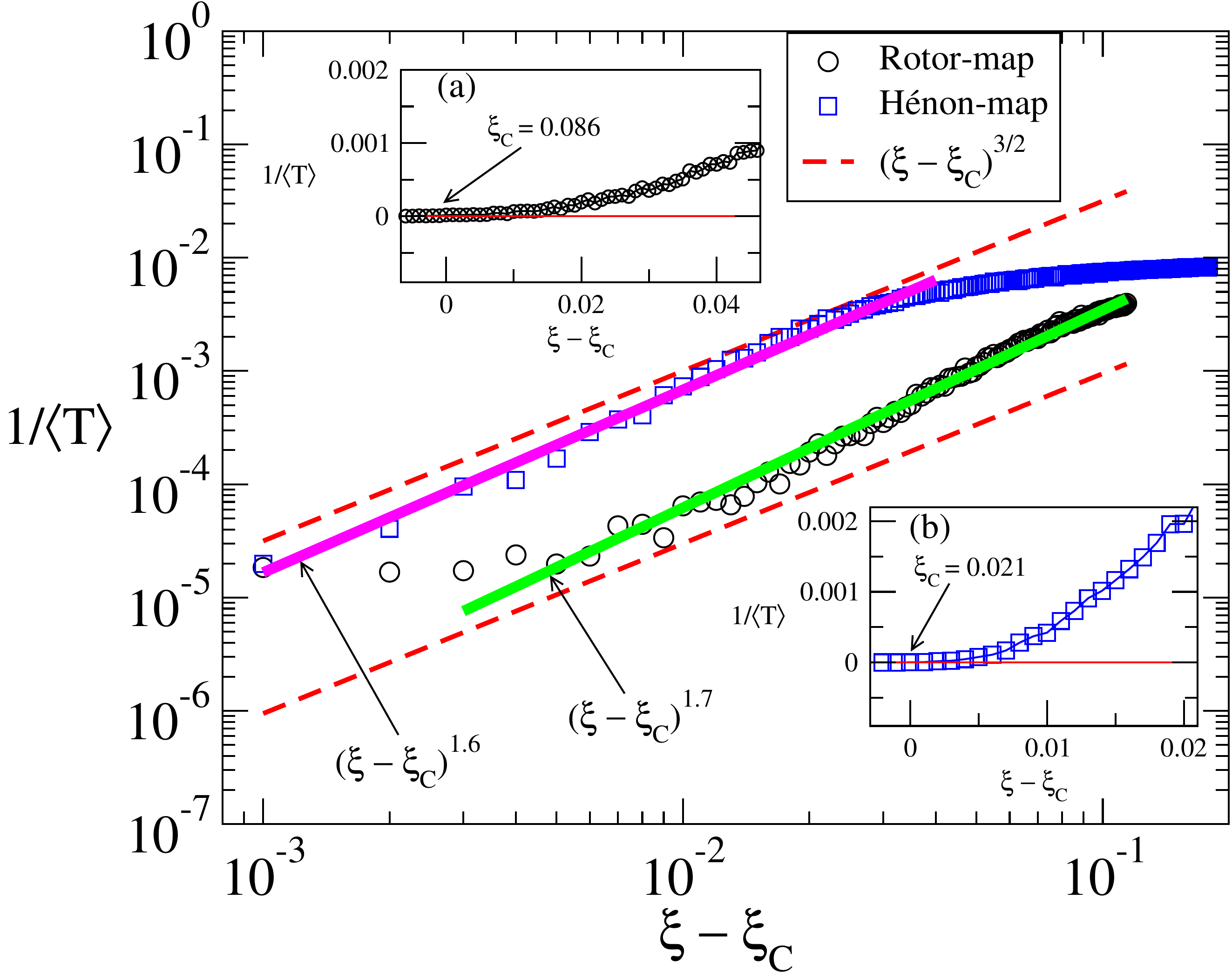}
\caption{(Colour online) The inverse of the mean escape time scaling
  with amplitude of noise for the Map (\ref{eq.single-rotor-map}) - black circles - and for the Map (\ref{eq.henon-map}) - blue squares. For each map, the values of mean escape time were obtained by iterating $10^{3}$
  random orbits for each value of $\xi$. The dashed lines show the expected scaling $(\xi -
  \xi_{c})^{3/2}$ and the thick continuous lines show the best fitting for the Rotor map,
  $\alpha \approx 1.7$, and for the H\'{e}non map,   $\alpha \approx 1.6$. In the insets, we show the transition. For $\xi
  < \xi_{c}$, we have $\langle T \rangle = \infty$, the random orbits
  do not escape, therefore $1/\langle T \rangle = 0$. For $\xi \geq \xi_{c}$, the escape time scales
  as Eq. (\ref{eq.meantime}). For the used parameters $\xi_{c} = 0.086 \pm 0.006$ for the Rotor map - inset (a) - and $\xi_{c} = 0.021 \pm 0.002$ for the H\'{e}non map - inset (b).}
\label{fig.escape}
\end{figure}

In order to check this prediction we numerically obtained the scaling of
the distribution of escaping times with amplitude of noise for two distinct $2$-dimensional systems.
The first perturbed systems we have chosen was the randomly perturbed single rotor map ~\cite{Zas78}, defined by
\begin{equation}
\label{eq.single-rotor-map}
F \left(\begin{array}{c}
x_{j} \\
y_{j} 
\end{array}\right) =
\left( \begin{array}{c} 
x_{j} + y_{j}(mod 2\pi )\\
(1 - \nu)y_{j} + 4 \sin(x_{j} + y_{j})
\end{array} \right) +
\left( \begin{array}{c} 
\varepsilon_{x_{j}}\\
\varepsilon_{y_{j}}
\end{array} \right),
\end{equation}
where $x \in [0,2\pi]$, and $y \in \mathbb{R}$, and $\nu$ represents the dissipation parameter.
As a second testing system, we have chosen the perturbed dissipative H\'{e}non map, in the form 
\begin{equation}
\label{eq.henon-map}
G \left(\begin{array}{c}
x_{j} \\
y_{j} 
\end{array}\right) =
\left( \begin{array}{c} 
1.06 x^{2}_{j} - (1 - \nu) y_{j}\\
                x_{j}
\end{array} \right) +
\left( \begin{array}{c} 
\varepsilon_{x_{j}}\\
\varepsilon_{y_{j}}
\end{array} \right),
\end{equation}
where, $x$ and $y$ are real numbers and again, $\nu$ represents the dissipation parameter. We used $\nu = 0.02$ 
for both maps, as for such value they present very rich dynamics~\cite{RMG09}. We also assumed, for the
purposes of this numerical experiment, the noise to be uniformly
distributed in each variable; but we stress that this is just a
numeric convenience. For each map, we
computed the time that random orbits took to escape from their respective main
attractors for a range of noise amplitudes. In each case, the mean escape time was
obtained for $10^{3}$ random orbits for each value of $\xi$. The results
are shown in Fig.~\ref{fig.escape}. For the parameter used here, we
obtained $\xi_{c} = 0.086 \pm 0.006$ for the perturbed Rotor map and $\xi_{c} = 0.021 \pm 0.002$ for the perturbed H\'{e}non map, what is shown in the insets. In both cases, we obtained
a good agreement between our simulations and the predictions of our
theory for a range of decades. An important remark is regarding the precision of the $\xi_{c}$. For the one dimensional case, for example, it has been proved that similar power laws in the unfolding parameters are in fact lower bounds for the average escape time scale~\cite{ZmH07}. Therefore, even from the numerical perspective, it is difficult to accurately estimate the value of $\xi_{c}$. 
Indeed, in our case we observe when increasing $\xi$ near $\xi_{c}$ some transient irregular bursts regime before the escape phenomena becomes robust. For a number of initial conditions, we have thus a distribution of values of critical noise around $\xi_{c}$, that is expected to become sharper as the number of initial conditions is increased. As a direct implication, the exponent obtained in our numerical simulations also varies within some range. For example, for the Rotor map, if we choose $\xi_{c} = 0.086 + 0.006$, we obtain $\alpha = 1.5$ and for $\xi_{c} = 0.086 - 0.006$, we have $\alpha = 2.0$. For the H\'{e}non map, we obtained $\alpha = 1.3$ for $\xi_{c} = 0.023$ and $\alpha = 2.1$ for $\xi_{c} = 0.019$.

Note also that, in principle, a much larger number of random orbits would be necessary for one to be able to observe a``perfect" power law. This is because most of the theoretical arguments used here, such as the convergence of time averages, are obtained in the asymptotic limit.
Furthermore, we notice that for ``large" values of $(\xi -
\xi_{c})$, meaning large holes, our simulated results differ
appreciably from our theoretical prediction. This is due to the fact
that $\mu_{c}(I_{\partial}) = \mu(I_{\partial})$ is valid only for
small leaks ~\cite{AlT08}. In addition, for large amplitude of noise,
the dynamics is totally dominated by the noise, and is not well
described as a small perturbation around the deterministic motion. We
note here that the exponential distribution reported in~\cite{KFG99}
for a particular case of bounded noise was obtained for values of
$\xi$ much greater than the critical amplitude $\xi_{c}$, which is an
outside the range of validity of our theory.

As last consideration, we want to call attention to the case where
noise is applied on bifurcating systems~\cite{SOG91, Ott02, DeG09}. In
our approach we consider $\eta$ to be increasing. In
Fig.~\ref{fig.condbound}, this corresponds to moving the parabola
upwards until it intersects the $\beta$ curve, when the
transition to escape takes place. We can easily see that we should expect
an equivalent transition to escape by moving $\beta$ instead, as a
result of changing a bifurcating parameter whilst keeping the noise
amplitude constant. Therefore, the exponent obtained for the case of
bifurcating systems is expected to be the same as ours, and this is indeed
the case~\cite{SOG91, Ott02, DeG09}.

In conclusion, we have shown that the problem of escaping orbits from
attractors due to the effect of bounded random noise can be thought as
a closed system with a hole. We identify the subset of the phase space
that is responsible for the escape and acts as a hole, and show that
the measure of this set determines the escape rate. We have shown that
there is a critical amplitude of noise in order to such escape
happens. When the amplitude is just above this critical value, we
derived a universal power-law relation of escape time with respect to
the amplitude of noise, in contrast with the case of Gaussian noise.

The authors a grateful to the anonymous referee for valuable suggestions to increase the value of this paper. 
AM and CG have been supported by the BBSRC, under grants BB-F00513X
and BB-G010722.


\end{document}